\documentstyle[ltwol,epsfig]{article}

\def\rar{\rightarrow}

\def\ap{$B \rar a_0\pi$}
\def\ar{$B \rar a_0\rho$}
\def\apr{$B \rar a_0\pi(\rho)$}
\def\bp{$B \rar b_1\pi$}

\def\apC{$B^{\pm} \rar a_0\pi$}

\def\aopopo{B^0 \rar a_0^0 \pi^0 }

\def\appm{B^0 \rar a_0^+\pi^-}
\def\ampp{B^0 \rar a_0^-\pi^+}
\def\appmb{\overline{B^0} \rar a_0^+\pi^-}
\def\amppb{\overline{B^0} \rar a_0^-\pi^+}

\def\appo{B^+ \rar a_0^+\pi^0}
\def\ampo{B^- \rar a_0^-\pi^0}
\def\aopm{B^- \rar a_0^0\pi^-}
\def\aopp{B^+ \rar a_0^0\pi^+}

\def\aopo{B^0 \rar a_0^0\pi^0}
\def\aopob{\overline{B^0} \rar a_0^0\pi^0}

\def\epp{$B^0 \rar \eta\pi\pi~$}

\def\eppc{$\eta\pi^+\pi^-~$}
\def\eppcd{$\eta\pi^-\pi^+~$}

\def\3pi{\pi\pi\pi}

\def\rp{$B \rar \rho\pi~$}
\def\pp{$B \rar \pi\pi~$}
\def\al{$\alpha$}
\def\ao{$a_0~$}
\def\eal{e^{+i\alpha}}
\def\emal{e^{-i\alpha}}
\def\babar{{\tt\it B{\small A}B{\small AR}}}
\def\BB{$B^0\overline{B^0}$}

\def\apm{A^{+-}}
\def\amp{A^{-+}}
\def\apmb{\overline{A}^{+-}}
\def\ampb{\overline{A}^{-+}}
\def\aoo{A^{00}}
\def\aoob{\overline{A}^{00}}

\def\Btappm{B^0(\Delta t) \rar a_0^+ \pi^-}
\def\Btampp{B^0(\Delta t) \rar a_0^- \pi^+}

\def\ie{{\it i.e.}~}
\def\eg{{\it e.g.}}
\def\vs{{\it vs}~}
\def\cf{{\it cf}~}

\begin{document}

\title{
\LARGE 
{\bf CP Violation and the Absence of Second Class Currents \\
in Charmless B Decays}
}
\author{Sandrine Laplace}

\address{Laboratoire de l'Acc\'el\'erateur Lin\'eaire, IN2P3-CNRS et Universit\'e Paris-Sud \\
BP 34, F-91898 Orsay Cedex, France \\E-mail: laplace@lal.in2p3.fr}   

\author{Vasia Shelkov}

\address{Lawrence Berkeley National Laboratory, 1, Cyclotron Road, Berkeley, CA 94720, USA \\
E-mail: VGShelkov@lbl.gov}

\twocolumn[\maketitle\abstracts{ 
The absence of second class currents together with the assumption of factorization for non-leptonic $B$ decays
provide new constraints on CP observables in decay $B \rar a_0(980)(\rar \eta\pi)\pi$. The kinematics of this
decay do not allow interference between the oppositely charged resonances in the Dalitz plot as 
in $B^0 \rar \rho(770)\pi~$. Nonetheless, under the assumption of factorization,
the \ap~two-body time-dependent isospin analysis leads to a more robust extraction of 
the angle \al~than in the \rp isospin-pentagon analysis. The absence of second class currents might 
lead to enhanced direct CP violation and/or allows for a test of 
some assumptions made in the \al~analysis in other decays like $B \to a_0 \rho$, $B \rar b_1(1235)\pi$, 
$B\to a_0 a_0$, $B\to \eta(\eta')\pi\pi$ and $B\to b_1 a_0$. The effects from non-factorizable 
contributions on the determination of \al~are estimated by means of a numerical study.}]
%
%

\section{Introduction}

The utility of the \ap~decay for measuring the angle \al~of the unitarity triangle by a 
time-dependent three-body Dalitz plot~\cite{Quinn2} or a two-body isospin~\cite{London,Quinn} analysis
has been emphasized by Dighe and Kim~\cite{Dighe}. 
It thus joins the list of channels like \pp and \rp allowing the extraction of $\alpha$.

These latter channels suffer from serious experimental limitations. The \pp decays
have low branching fractions~\cite{CLEO,Babar} and measuring the $\pi^0\pi^0$ final state 
is an experimental challenge. 
The branching ratio of the \rp decay is larger~\cite{CLEO,Babar} but this channel
suffers from combinatorial background due to the presence of a $\pi^0$ and contamination from higher
excitations~\cite{Sophie}, which complicate the time-dependent Dalitz-plot analysis.
The \ap~decay~\cite{BABARa0pi} has some advantages from the experimental point of view,
as pointed out by Dighe and Kim~\cite{Dighe}, since it is easier to reconstruct a $\eta$
than a $\pi^0$ (due the higher energies of the final state photons) 
and  since the width of the \ao is narrower (around 60 MeV~\cite{PDG}) than the width of the $\rho$ 
(150 MeV~\cite{PDG}). 
These properties help to reduce the combinatorial background, and should thus provide a cleaner 
signal sample than for the \rp mode.

However, the interference pattern, which is effective in \rp, is kinematically excluded in \ap.
There is simply no overlap between the $\appm$($\rar$ \eppc) and $\ampp$($\rar$ \eppcd) bands in the Dalitz plot,
which provides the main source of interference in the \rp channel..

Focusing on the decays \ap~and \ar, we show in this paper that their analysis as
two-body decays, because of the absence of second class currents\footnote{ This 
was first pointed out to us by J. Charles in a private communication.}, and under the
assumption of naive factorization, leads to a more robust\footnote{ The 
analysis is more robust in the sense that there are either more degrees of freedom 
or less unknowns in the fit extracting \al, which makes the fit more stable.}
determination of the angle \al, than the original isospin-pentagon analysis proposed by Lipkin, Nir, Quinn and 
Snyder~\cite{Quinn} for \rp and applied to \ap~by Dighe and Kim~\cite{Dighe}. 
The effects of non-factorizable contributions are studied thanks to a likelihood analysis.

The time-dependent two-body \apr~analyses proceed through seven to nine-parameter
fits depending on whether or not the charged modes are considered.
When statistics are limited, simpler four-parameter fits can be performed for 
\apr~decays by using one theoretical prediction of an amplitude (or a ratio of two
of them)~\cite{Jerome}.

Moreover, as advocated in Section \ref{par:dircp}, the elimination of leading tree contribution due to the 
suppression of second class currents may give rise to enhanced direct CP violation in the decay \ap, as well
as \bp~ and $B\to \eta(\eta')\pi\pi$.

Finally, the \ap, \ar, $B\to a_0 a_0$, $B\to b_1 b_1$ and $B\to a_0 b_1$ decays provide a means for an evaluation of the 
non-factorizable contributions.

\section{The absence of Second Class Currents in some Non-Leptonic $B$ Decays}
\label{sec:SCC}

In tree diagrams contributing to non-leptonic $B$ decays, part of the hadronic system is produced via
coupling of the virtual $W$ to the quark current. Charmless final states with zero net strangeness proceed via 
the $W^+\to u\bar{d}$ coupling, with rates proportional to the CKM matrix element $|V_{ud}|^2$. 

In the naive factorization, the color singlet pair of quarks hadronizes independently 
of the rest of the $B$ decay. This implies that there is no re-scattering (or final state interaction)
between the hadrons coming from the $W$ and the other hadrons of the final state. Under this
assumption, the production of hadrons resulting from the coupling of quarks to the virtual $W^\pm$ abide 
by the same rules as semi-leptonic $\tau$ decays. We recall some of the relevant properties in the following.

The vector part of the weak current $\bar{u}\gamma_{\mu}(1-\gamma_5)d$ has even
$G$-parity, whereas the axial part has odd $G$-parity. It follows that a virtual $W^+$ 
decaying to $u\bar{d}$ produces states with an even $G$-parity and natural
spin-parity $(0^+, 1^-, ...)$, or with an odd $G$-parity and unnatural spin-parity
$(0^-, 1^+, ...)$. Decays with opposite combinations of $G$- and spin-parity are called
second class currents, and are forbidden in the Standard Model up to isospin violations.
This is the case for the $a_0$ which has $G=-1$ and $J^P=0^+$, and the $b_1$ which has $G=+1$ and $J^P=1^+$.
Experimental limits on second class currents are obtained, \eg, from the measurement of 
$\tau^+\rar\eta \pi^+\nu_{\tau}$ branching fraction for which the present limit reads $1.4 \times 10^{-4}$ 
at $95 \% ~{\rm CL}$~\cite{PDG}.

States with $J^P = 0^+$ are also forbidden by the conservation of the vector current, 
independently of their $G$-parity, up to isospin violating corrections. 
Therefore the $W\rar a_0$ decay is doubly-suppressed.

Wether the potential non-factorizable contributions are small corrections
or as large as the factorizable terms is a controversial question. non-factorizable contributions 
effects on the \al~analysis are described in section ~\ref{sec:study}.

In addition to assuming naive factorization, contributions from annihilation and exchange 
diagrams are neglected since they are expected to be suppressed by helicity conservation 
and by the quantity $f_B/m_B$~\footnote{This arises from dimensional arguments.},
where $f_B$ is the decay constant of the $B$.

Thus, under these assumption, the absence of second class currents leads to the suppression 
of tree diagrams in which the $a_0$ ($b_1$) and the virtual $W$ have the same charge.

Experimental tests of the factorization assumption and measurements of the non-factorizable
terms for the decays treated in this paper are proposed in Sec.~\ref{sec:ampl}
and ~\ref{sec:aa-bb}.

\section{Extracting $\alpha$ from \ap~and \ar~Decays}

This section aims at showing the consequences of the absence of second class currents
in the extraction of \al~ in the \ap~and \ar~ decays.
The phase-space analyses of \ap~ and \ar~ are not as powerful as for \rp, since the interferences
between the different resonances are weak (\cf Sec.~\ref{three-body} and~\ref{four-body}).

The emphasis is put on the \apr~ time dependent two-body analysis, which can be performed separately for 
$B \to a_0\pi$ and $B \to a_0\rho$. In effect, one could use both modes in a combined fit, 
hence reducing the number of mirror solutions for the angle \al~ (\cf Sec.~\ref{par:mirror}).

On the one hand, the branching ratio of  $B^0 \to a_0\rho$ is expected to be larger then that for
\ap, just as $BR(\tau \rar \rho \nu ) > BR(\tau \rar \pi \nu)$. 
On the other hand, decays involving a charged $\rho$ ($\rho^{\pm}\to\pi^{\pm}\pi^0$) 
require the reconstruction of an additional $\pi^0$. Finally, in contrast to \ap, the time-dependence 
of $B^0 \to a_0^0\rho^0$ is measurable due to the charged products of the $\rho^0\to\pi^+\pi^-$.

Naive factorization is assumed throughout this section.

\subsection{Tree and Penguin Contributions and Consequences of the Absence of Second Class Currents}
\label{sec:ampl}

In processes involving $u\bar{u}d$ non-spectator quarks, the decay amplitude can be expressed 
in terms of the tree ($T$) and $u$-, $c$- and $t$-penguin ($P^u, P^c, P^t$) contributions
(where the CKM matrix elements have been explicitely factorized out):
{\small
\begin{eqnarray}
A (u\bar{u}d) & = V_{tb}V_{td}^* P^t + V_{cb}V_{cd}^* P^c + V_{ub}V_{ud}^* (T + P^u)~, \nonumber \\
              & = V_{tb}V_{td}^* (P^t - P^c) + V_{ub}V_{ud}^* (T + P^u - P^c)~. \label{eq:contr2}
\end{eqnarray}}
The second line is obtained by using the unitarity relation $V_{ub}V_{ud}^*+V_{cb}V_{cd}^*+V_{tb}V_{td}^*=0$.
The amplitude is thus the sum of two terms depending on the weak phases $\beta$ (from $V_{td}^*$) and $- \gamma$ (from $V_{ub}$).
We will neglect the contributions from $P^u$ and $P^c$ and propose a test of this assumption later in this section. Therefore, 
the remaining $t$-penguin provides $\beta$, whereas $\gamma$ is only invoked by the tree amplitude.
We will denote these two contributions $T$ and $P$ in the following, where $P$ is restricted to the $t$-penguin 
contribution only.

The $B^0 \rar a_0^i \pi^j/\rho^j$ (with $i,j = 0,+,-$) decay amplitudes $A^{ij}$ can thus be expressed in terms of
tree ($T^{ij}$) and penguin ($P^{ij}$) contributions and the weak phase \al. For example, the amplitudes for 
the \ap~ decay read:
{\small
\begin{eqnarray}
A(B^0 \rar a_0^+ \pi^-) & =  \apm  & =  \emal T^{+-} + P^{+-}~,
\label{eq:PandT1} \\
A(B^0 \rar a_0^- \pi^+) & =  \amp  & =  \emal T^{-+} + P^{-+}~, \\
A(B^0 \rar a_0^0 \pi^0) & =  \aoo  & =  \emal T^{00} + P^{00}~, \\
A(\overline{B^0} \rar a_0^+ \pi^-) & =  \apmb & =  \eal T^{-+} + P^{-+}~, \\
A(\overline{B^0} \rar a_0^- \pi^+) & =  \ampb & =  \eal T^{+-} + P^{+-}~, \\
A(\overline{B^0} \rar a_0^0 \pi^0) & =  \aoob & =  \eal T^{00} + P^{00}~. 
\label{eq:PandT6}
\end{eqnarray}
}where the $q/p$ mixing parameter~\cite{BabarBook} has been absorbed in the $\bar{A}$ amplitudes, leading
to the explicit presence of the angle $\alpha$.
The $T^{+-}$ amplitude comes from the $W^+ \to a_0^+$ transition, and is 
suppressed as a Second Class Current Forbidden Tree (\mbox{SCCFT}). 
Therefore, the $A(B^0 \rar a_0^+ \pi^-/\rho^-)$ and $A(\overline{B^0} \rar a_0^- \pi^+/\rho^+)$ 
amplitudes are pure penguin transitions, and cannot display direct CP violation:
{\small
\begin{eqnarray}
A(B^0 \rar a_0^+ \pi^-) & = & \apm  =  P^{+-} ~,
\label{eq:PandT1b} \\
A(\overline{B^0} \rar a_0^- \pi^+) & = & \ampb =  P^{+-} ~,
\label{eq:PandT6b}
\end{eqnarray}}
and therefore
{\small
\begin{equation}
A(B^0 \rar a_0^+ \pi^-) = A(\overline{B^0} \rar a_0^- \pi^+)~.
\label{eq:eq} 
\end{equation}}
Equality~(\ref{eq:eq}) follows from the absence of the $V_{ub}V_{ud}^*$ term in Eq.~\ref{eq:contr2}. This, 
in turn, resulted from SCCFT killing the tree contribution and our assertion that $(P^u - P^c)$ could be ignored.
Both are open to challenge. Failure of factorization could introduce a tree contribution. The non-$t$ penguins
might not be small. Thus the $V_{ub}V_{ud}^*$ term cannot be completely excluded, although it follows from 
commonly made approximations. In addition, even if Eq.~(\ref{eq:eq}) is verified experimentally, that would
not prove that the $V_{ub}V_{ud}^*$ term is absent. For Eq.~(\ref{eq:eq}) to be violated, there must be
differing strong phases from the $V_{tb}V_{td}^*$ and $V_{ub}V_{ud}^*$ amplitudes and little can be said
with confidence about such strong phases a priori. Nonthless, experimental confirmation of Eq.~(\ref{eq:eq})
would give circumstantial evidence in favor of the assumptions made here.

\subsection{The \ap~Three-Body Analysis {\it \`{a} la $\rho\pi$}}
\label{three-body}

Dighe and Kim~\cite{Dighe} have proposed to extract \al~ from the \ap~decay using both two-body
isospin and three-body Dalitz plot analyses.

\begin{figure}[h]
\centerline{\psfig{figure=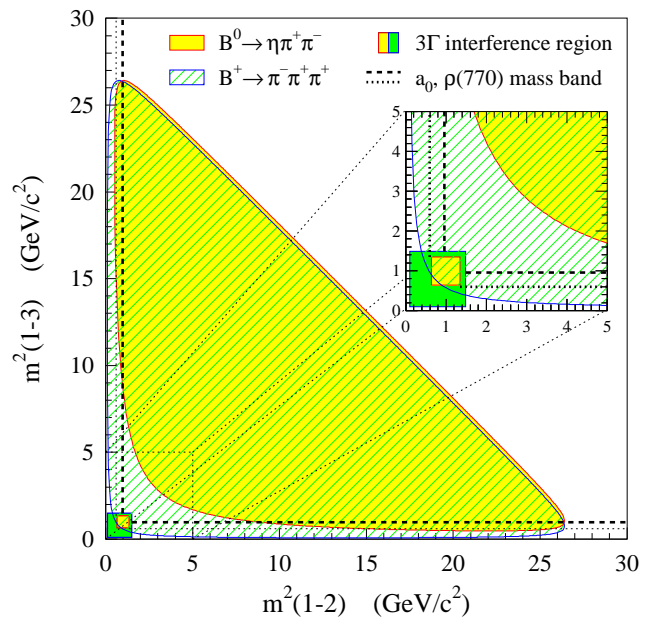,width=8.5cm}}
\caption[.] {\label{fig:Dalitz}
{\small 
Dalitz plot kinematic boundaries for the \epp and $B^0 \rar \pi^0\pi^+\pi^-$ decays.
The dashed (dotted) line shows the $a_0$ ($\rho$) mass band. The $3\Gamma$ interference regions
for the $a_0$ (light shade) and $\rho$ (dark shade) resonances are also drawn, the region for the $a_0$
lying outside the allowed boundary of the $\eta\pi^+\pi^-$ Dalitz plot.}}
\end{figure}

The Dalitz-plot analysis fails because of the small interference between the 
oppositely-charged $a_0^{\pm}$, as shown in Fig.\ref{fig:Dalitz}.
Since most of the interference occur when the two resonance bands intersect, the regions covering
three times the width (called ``$3\Gamma$ interference region'') are indicated for the $a_0$ and $\rho$
resonances. Kinematic boundaries for $B^0 \rar \eta\pi^+\pi^-$ and $B^+ \rar \pi^-\pi^+\pi^+$ are also drawn.
The shape of the boundary in the left-hand bottom corner of the \ap~Dalitz plot is  
determined by the $\eta$ mass, which limits the available phase-space.
In contrast to $\rho$ in the $B^+ \rar \pi^-\pi^+\pi^+$ decay, 
the $a_0$ mass and width are too small to allow strong interferences within the kinematic limits of the Dalitz plot.

Interference can still occur far away from the $3\Gamma$ intersection region, but it is less than
$1\%$ in the case of \ap\footnote{ \cf Sec.~\ref{four-body} for the description of a method on how to compute the strength 
of the interference.} and occur in the badly-known tails of the $a_0$ resonance.

Therefore, the Dalitz-plot analysis for \ap~ is not of interest.

\subsection{The \ar~Four-Body Analysis}
\label{four-body}

The modes $B^0\to a_0^+\rho^-$, $B^0\to a_0^-\rho^+$ and $B^0\to a_0^0\rho^0$ decay into the common 
four-body final state $\eta \pi^+ \pi^- \pi^0$. If interference between $a_0$'s and $\rho$'s is strong enough, 
one could perform a similar time and phase-space dependent analysis as for $B\rar \rho\pi$. 

To quantify the strength of the interference, the following parameter~\cite{Sophie} can be evaluated:
\begin{equation}
\epsilon=| \sum_{i=1}^{3}f_i |^2 \Big/ \sum_{i=1}^{3}|f_i|^2 - 1~,
\end{equation}
where $f_1=f(a_0^+)f(\rho^-)\cos\theta$, $f_2=f(a_0^-)f(\rho^+)\cos\theta$, and 
$f_3=f(a_0^0)f(\rho^0)\cos\theta$ are the products of the $a_0$ and $\rho$ Breit-Wigners, taking into account 
the distribution of the helicity angle $\theta$ (defined as the angle between the $\rho$ decay axis in the
$\rho$ rest frame and the direction of the $\rho$ in the laboratory frame).
Using simple relativistic Breit-Wigner parameterizations for the $\rho$ and the $a_0$ resonances\footnote{ The 
$a_0$ mass parameterization is complicated by the $KK$-production threshold~\cite{PDG}, and is
not well-known. Using a simple Breit-Wigner is a rough approximation.}, 
the $\epsilon$ parameter distribution is computed using \ap~Monte Carlo events. 
The mean value of $|\epsilon|$ is equal to $\sim 10\%$, corresponding to roughly half of what 
is observed in \rp~\cite{Sophie}. Therefore the \ar~decay provides only limited interference effects.

Additional complication of having to reconstruct an extra neutral particle makes this channel
less accessible than $B\to \rho\pi$.  Nevertheless, the time and phase-space dependent analysis 
of the $B^0\to a_0\rho$ decay provides an independent and complementary way of measuring the 
angle $\alpha$ without any ambiguities.

\subsection{The \apr~Two-Body Time-Dependent Analyses}
\label{sec:TwoBody}

Since the \epp three-body final state does not exhibit interference in the Dalitz plot, 
one is led to a two-body analysis, \ie where $\appm$ and $\ampp$ decays are considered as 
two-body final states. The analysis can be applied to \ar~as well.

The time-dependent amplitudes for the two-body decays $\Btappm$ and $\Btampp$ 
(as well as for the CP-eigenstate $B^0 \rar a_0^0\rho^0$) read:
{\small 
\begin{eqnarray}
&& A(\Btappm) \propto  e^{-\frac{\Gamma |\Delta t|}{2}} \nonumber\\[4pt]
&&{}  \times \left[ \cos(\frac{\Delta m\Delta t}{2}) \apm + i~\sin(\frac{\Delta m\Delta t}{2}) \apmb \right]
\label{eq:Btappm} \\[2pt]
&&{} A(\Btampp) =  e^{-\frac{\Gamma |\Delta t|}{2}} \nonumber\\[4pt]
&&{}  \times \left[ \cos(\frac{\Delta m\Delta t}{2}) \amp + i~\sin(\frac{\Delta m\Delta t}{2}) \ampb \right]~,
\label{eq:Btampp}
\end{eqnarray}}
where the cosine and sine terms describe the \BB~flavor mixing, and $\Delta t$ is the 
difference of decay time between the two $B$ mesons produced at the $\Upsilon(4S)$ resonance
in an asymmetric B factory. The $\apm$, $\amp$, $\apmb$ and $\ampb$ amplitudes are defined in 
Eqs.~(\ref{eq:PandT1})-(\ref{eq:PandT6b}).

The time-dependent decay rate is obtained by squaring Eqs.~(\ref{eq:Btappm}) and (\ref{eq:Btampp}), 
which leads to terms proportional to $\sin^2(\Delta m\Delta t/2)$, $\cos^2(\Delta m\Delta t/2)$ 
and $\sin(\Delta m\Delta t)$:
{\small
\begin{eqnarray}
&& \Gamma(B^0(\Delta t) \rar a_0^{\pm}\pi^{\mp}) \propto e^{-\Gamma |\Delta t|} \nonumber\\[4pt]
&&{} \times \left[ {\cal A}^{\pm}_{1} \sin^2(\frac{\Delta m\Delta t}{2}) + 
{\cal A}^{\pm}_{2} \cos^2(\frac{\Delta m\Delta t}{2}) + 
 {\cal A}^{\pm}_{3} \sin(\Delta m\Delta t) \right] \nonumber\\[2pt]
&&{} \propto  e^{-\Gamma |\Delta t|} \left[ {\cal A'}^{\pm}_{1} + 
{\cal A'}^{\pm}_{2} \cos(\Delta m\Delta t) + 
 {\cal A}^{\pm}_{3} \sin(\Delta m\Delta t) \right]~,
\label{eq:Rate}
\end{eqnarray}}
where the ${\cal A}^{\pm}_{1,2,3}$, ${\cal A'}^{\pm}_{1,2}$ terms are combinations of the 
$a_0^{\pm}\pi^{\mp}$ amplitudes.

Therefore, each time-dependent $\appm(\rho^-)$, $\ampp(\rho^+)$ and $B^0 \rar a_0^0(\rho^0)$ measurement provides 
three observables: ${\cal A'}_1$, ${\cal A'}_2$ and ${\cal A}_3$. 

The measurement of the branching ratios for charged $B$ decays \apC$(\rho)$ and/or for the neutral final 
state $\aopopo(\rho^0)$ each provides one observable.
Using isospin invariance~\cite{London,Quinn,BabarBook}, 
one can link the penguin and tree contributions from neutral and charged B decays, which provides
the missing pieces for the extraction of \al:
{\small
\begin{eqnarray}
\sqrt{2}\left[ T^{+0} + T^{0+} \right] & = & T^{+-} + T^{-+} + 2 T^{00}~, \label{isosp1} \\
P^{00} & = & - \frac{1}{2}(P^{+-} + P^{-+}) ~, \label{isosp3}\\
P^{+0} & = & \frac{1}{\sqrt{2}}(P^{+-} - P^{-+}) ~, \label{isosp4} \\
P^{0+} & = & - \frac{1}{\sqrt{2}}(P^{+-} - P^{-+})~.  \label{isosp5}
\end{eqnarray}}

Table \ref{AllDecay} gives a comparison of the number of observables and unknowns for
\ap, \ar, \rp and \pp analyses. Three analyses steps are described: in the upper part of the table, only 
charged final states of neutral $B$ decays are used. In the middle part, neutral final states
of neutral $B$ decays are added. In the lower part, both neutral and charged $B$ decays are taken
into account. Available isospin relations are indicated at each analyses stage.
\begin{table*}[t]
\begin{center}
\begin{tabular}{|c|c|c|c|c|c|c|c|c|c|}
\hline
  Channel & Contributing &\multicolumn{2}{c|}{$a_0\pi$}
  &\multicolumn{2}{c|}{$a_0\rho$}
  &\multicolumn{2}{c|}{$\rho\pi$}&\multicolumn{2}{c|}{$\pi\pi$} \\ 
\cline{3-10}
  Ex: \ap & T \& P Amplitudes
  & ${\cal O}$   &  ${\cal U}$  &  ${\cal O}$    & ${\cal U}$  
  &  ${\cal O}$    &  ${\cal U}$   & ${\cal O}$   & ${\cal U}$ \\ 
  \hline &&&&&&&&&\\[-8pt]
$\matrix{ \appm \\ \appmb}  $ & 
  $\matrix{\emal T^{+-}+ P^{+-} \\\eal T^{-+}+ P^{-+}}$  
  &3$_t$& $\matrix{5 \\ 4}$ & 3$_t$ & $\matrix{5 \\ 4}$ & 3$_t$ 
  & $\matrix{5 \\ 4}$  & 3$_t$ &  $\matrix{5 \\ $-$}$ \\[6pt]
$\matrix{ \ampp \\ \amppb} $ & 
  $\matrix{ \emal T^{-+}+ P^{-+} \\ \eal T^{+-}+ P^{+-}}$ 
  & 3$_t$ & - & 3$_t$ & - & 3$_t$ & -  & - & - \\[8pt]
 \cline{1-2}
  \multicolumn{2}{|c|} 
  {Overall norm. \& phase} &$-$1 & $-$2& $-$1  &
  $-$2& $-$1 & $-$2 & $-$1 & $-$2 \\[1pt]
\cline{1-2} \multicolumn{2}{|c|}{\mbox{SCCFT} ($T^{+-}=0$)} 
  &$-$1 & $-$2 & $-$1  & $-$2 &  &  &  &  \\ \hline\hline 
\multicolumn{2}{|c|}{Total using only $B^0$'s}                             
  &\multicolumn{2}{c|}{4 \vs5}  &\multicolumn{2}{c|}{4 \vs5} 
  &\multicolumn{2}{c|}{5 \vs7}  &\multicolumn{2}{c|}{2 \vs3} \\ 
  \hline\hline &&&&&&&&&\\[-8pt]
$\matrix{ \aopo \\ \aopob} $ &
  $\matrix{\emal T^{00}+ P^{00} \\ \eal T^{00}+ P^{00}}$ &
  $\matrix{1_i \\ 1_i}$ &  $\matrix{4 \\ $-$}$ & 3$_t$ 
  & $\matrix{4 \\ $-$}$ & 3$_t$ & $\matrix{4 \\ $-$}$ &
  $\matrix{1_i \\ 1_i}$ & $\matrix{4 \\ $-$}$ \\[8pt]
\cline{1-2}\multicolumn{2}{|c|} {Isospin relation (\ref{isosp1}) }
  &  & $-$2&  & $-$2&  & $-$2  &  & $-$2 \\ \hline\hline
\multicolumn{2}{|c|} {Total adding neutral final state}                        
  &\multicolumn{2}{c|}{6 \vs7} &\multicolumn{2}{c|}{\bf 7 \vs7} 
  &\multicolumn{2}{c|}{8 \vs9} &\multicolumn{2}{c|}{4 \vs5}
  \\  \hline\hline  &&&&&&&&&\\[-8pt]
$\matrix{ \appo \\[0.8pt] \aopp} $ &
  $\matrix{\emal T^{+0}+ P^{+0} \\ \emal T^{0+}+ P^{0+}}$ &
  $\matrix{1_i \\ 1_i}$ & $\matrix{4 \\ 4}$ 
  & $\matrix{1_i \\ 1_i}$ & $\matrix{4 \\ 4}$ & 
  $\matrix{1_i \\ 1_i}$ & $\matrix{4 \\ 4}$ &
  $\matrix{1_i \\ $-$}$ & $\matrix{4 \\ $-$}$ \\ [9pt]
$\matrix{  \ampo \\[0.8pt] \aopm} $ &
  $\matrix{ \eal T^{+0}+ P^{+0} \\\eal T^{0+}+ P^{0+}}$ &
  $\matrix{1_i \\ 1_i}$ & - & $\matrix{1_i \\ 1_i}$ & - & 
  $\matrix{1_i \\ 1_i}$ & - & $\matrix{1_i \\ $-$}$ & - \\[8pt]
\cline{1-2}\multicolumn{2}{|c|}
  {Isospin relations (\ref{isosp3})-(\ref{isosp4}) }
  &  & $-6$&  & $-6$&  & $-6$  & & $-4$  \\  \hline\hline
\multicolumn{2}{|c|}{Total adding charged $B$'s} 
  &\multicolumn{2}{c|} {\bf 10 \vs 9}  &\multicolumn{2}{c|}{\bf 11 \vs9} 
  &\multicolumn{2}{c|}{\bf 12 \vs11} &\multicolumn{2}{c|}{\bf 6 \vs5} \\ \hline
\end{tabular}
\caption{\label{AllDecay}
{\small 
Number of observables (${\cal O}$) and unknowns (${\cal U}$) involved in the 
\ap~and \ar~analyses compared to the \rp and \pp analyses. \underline{Upper part:}
charged final states of neutral $B$ decays. \underline{Middle part:} neutral final states of neutral $B$ decays.
\underline{Lower part:} charged $B$ decays.
The time-dependence of neutral $B$ decays yields three observables (\cf Eq.~(\ref{eq:Rate}))
indicated with a ``{\bf t}'' subscript, whereas the ``{\bf i}'' subscript corresponds to time-integrated 
measurements (yielding a single observable). The fact that one can exchange the two pions in the
\pp final state removes half of the contribution to the number of observables and unknowns.
An overall normalization and phase are subtracted from the number of unknowns, and a normalization
is subtracted from the number of observables.
The \mbox{SCCFT} argument applies to the \ap~and \ar~ channels, removing one observable (because two of them 
turn out to measure the same quantity) and two unknowns.
The number of constraints coming from isospin relations is given when available.
The total number of observables \vs unknowns is indicated with {\bf bold} characters when the
fit is constrained.}}
\end{center}
\end{table*}

The leading contribution to $B^0 \rar a_0^+\pi^-$ , the $T^{+-}$ tree, is suppressed 
by \mbox{SCCFT}. 
One of the two contributions to the color-suppressed $T^{00}$ amplitude is removed by the same 
\mbox{SCCFT} argument\footnote{ This is because this contribution to the $T^{00}$ amplitude is the Fierz-transform 
of $T^{+-}$, therefore the same properties than for $T^{+-}$ hold.}, but the other contribution remains.
The leading contribution to the $T^{+0}$ amplitude is removed by \mbox{SCCFT}, but a color-suppressed
contribution remains.

The number of unknowns is given by the sum of tree and penguin complex amplitudes involved at each
analyses stage, plus the angle \al. One unphysical overall phase and one irrelevant overall normalization
constant are subtracted from the total.

The number of observables available from a time dependent measurement is 
three (\cf Eq.~(\ref{eq:Rate})), and one for the time integrated measurement. 
The overall normalization is subtracted from the sum of observables.

Using only the charged final states of the neutral $B$ decays does not provide enough observables
to constrain \al~ in any of the four analyses considered. Nevertheless, using a single theoretical prediction
for an amplitude (or a ratio of amplitudes) in four-parameter \apr~and two-parameter \pp fits would 
be enough to extract the value of $\alpha$. Such a model-dependent approach can be performed with
low statistics.

Adding the neutral final states does not further constrain the fits, either for \ap,
or for \rp, \pp. In contrast, the \ar~analysis 
does improve, since time-dependence is observable and \mbox{SCCFT} holds, 
though the fit is only barely constrained (seven observables \vs~seven unknowns).

Adding charged $B$ decays in the analyses allows all four fits to converge, but with differing
robustness: whereas the \rp~two-body analysis consists of an eleven-parameter fit with one extra constraint, 
in the \ap~analysis, \mbox{SCCFT} decreases the number of parameters to nine, with one 
extra constraint. As a consequence, \mbox{SCCFT} makes the \ap~analysis more robust. The \ar~analysis invokes a 
nine-parameter fit with two extra constraints, and finally, being a CP eigenstate, the \pp analysis is the 
simplest and is performed via a five-parameter fit.

Similarly to the \pp analysis, the requirement to measure the $\aopo$ branching ratio makes the \ap~analysis 
far more difficult.

\subsection{Mirror Solutions}
\label{par:mirror}

 CP violation in channels that benefit from \mbox{SCCFT} arises from interference between tree and penguin diagrams. 
Consequently, one measures $\alpha$-dependent terms like $\sin\alpha$ and $\cos\alpha$.
This is different from the \rp analysis where tree-tree interferences dominate and 
result in terms like $\sin 2 \alpha$ and $\cos 2 \alpha$.
 
 The extraction of $\alpha$ via \ap~is done through terms like $\sin(\alpha+\delta)$ 
and $\sin(\alpha-\delta)$, where $\delta$ is a strong phases difference. It thus leads
to multiple mirror solutions for $\alpha$ in the interval $[0,\pi]$, 
as in the two-body analyses of \pp and \rp.

In general, the number of mirror solutions depends on the type of analysis (\eg, one solution 
for the time-dependent Dalitz plot approach in \rp, but eight solutions for the \pp isospin analysis). 
To overcome this difficulty, the angle $\alpha$ has to be measured independently in various channels. 

\subsection{Possible Enhancement of Direct CP Violation}
\label{par:dircp}

 Even though direct CP violation is most frequently searched for with charged $B$ mesons,
neutral $B$ decays can also be used to look for possible asymmetries in untagged sample\footnote{ Untagged 
events should enter the \al~analysis as well.}:
{\small
\begin{eqnarray}
{\cal B}(B^0 \to a_0^+\pi^-) & + & {\cal B}(\overline{B^0} \to a_0^+\pi^-) \neq \\
{\cal B}(B^0 \to a_0^-\pi^+) & + & {\cal B}(\overline{B^0} \to a_0^-\pi^+)~,
\end{eqnarray}}
as well as in the tagged sample:
{\small
\begin{equation}
{\cal B}(\overline{B^0} \to a_0^+\pi^-) \neq {\cal B}(B^0 \to a_0^-\pi^+)~.
\end{equation}}
Indeed, the suppression of the leading tree due to \mbox{SCCFT} may enhance direct CP violation, 
provided that the remaining $T^{-+}$ and $P^{-+}$ are of comparable magnitude.
Similarly, in the charged $B$ decays, the interference of the remaining color-suppressed tree ($T^{+0}$) 
and the non-dominant tree ($T^{0+}$) with penguin contributions may enhance direct 
CP violating effects.

In contrast to the extraction of \al, the enhancement of direct CP violation in the \ap~ channel
does not depend on the hypotheses made in Sec.~\ref{sec:SCC} (factorization and neglecting $u$- and
$c$-penguin contributions), since a failure of the latter would not re-establish the hierarchy between
dominant trees and penguins. The possible enhancement of direct CP violation only stems from the absence of second class 
currents which is experimentally established.

\section{Likelihood Analysis}
\label{sec:study}

To assess the sensitivity to \al, and to probe the effects of non-factorizable contributions,
the four time-dependent (Eq.~\ref{eq:Rate}) and six time-independent measurements 
(see table \ref{AllDecay}) are implemented in a likelihood analysis. 
For this toy experiment, tree and penguin amplitudes are assumed to 
be the same as for the \rp mode (apart for the SCCFT tree $T^{+-}$)
as in Ref.~\cite{BabarBook}, and \al~is taken to be equal to 1.35 rad. 
These values determine in particular the position of the mirror solutions. 
The analysis assumes a total of $1500$ events,
which roughly corresponds to an integrated luminosity of $500\ fb^{-1}$,
with a typical selection efficiency of $10\%$.
 
The SCCFT effect is described by a factor $f$ applied to the $T^{+-}$ contribution of
Ref.~\cite{BabarBook}:
\begin{equation}
T^{+-}_{\rm a_0\pi} = f \times T^{+-}_{\rm \rho \pi},
\end{equation}
where $f=0$ corresponds to naive factorization and a non-zero $f$ value mimics non-factorizable contributions.
The analysis of the events generated with this set of amplitudes (where $f$ varies, \eg, from $0$ to $20 \%$) 
relies on the factorization hypothesis, \ie, $f=0$.

\begin{figure}[h]
\centerline{\psfig{figure=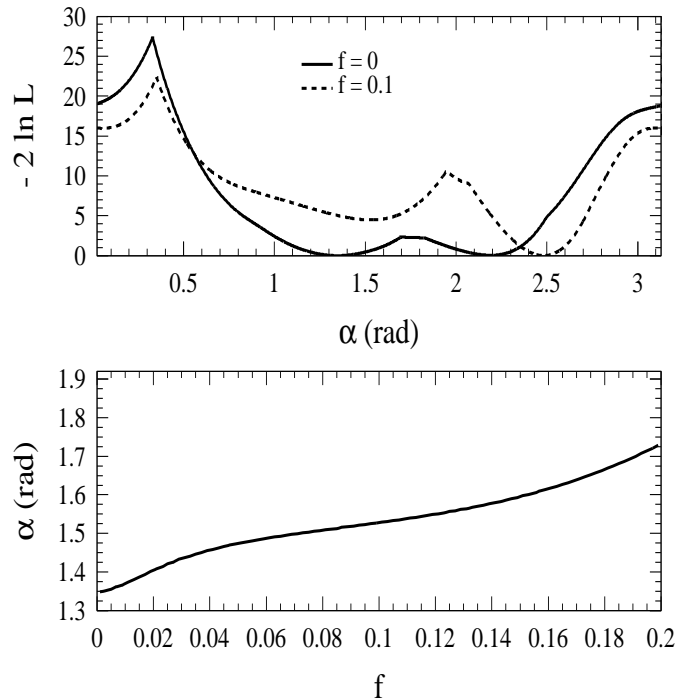,width=10cm,height=10cm}}
\caption[.] {\label{fig:scan}
{\small \underline {Upper plot}:
$chi^2(\alpha)=-2\ln{\cal L}$ functions for different values of $f$, the factor applied
to the \rp $T^{+-}$ tree to mimic non-factorizable contributions to \ap $f=0$ corresponds
to naive factorization. 
A mirror solution nearly degenerate with the main mininum is observed at $\alpha\simeq 2.2$ rad.
The minimum of the mirror solution becomes more pronounced as $f$ increases. For large
values of $f$, it becomes a global minimum.
\underline{Lower plot}: Position of one of the local minima (the one located on the 
true value of \al~for $f=0$.) as a function of $f$. 
}}
\end{figure}   

Figure ~\ref{fig:scan} shows the effects of non-factorizable contributions on the likelihood fit.
The upper plot displays $\chi^2=-2\ln{\cal L}$ functions for $f=0$ and $f=0.1$.  
By construction, for $f=0$, a minimum is located at the true value of \al:
this is because an analytical expression is used for the likelihood.
A pronounced mirror solution is visible for $\alpha\simeq 2.2$ rad. 
For increasing values of $f$, this mirror solution deepens and evolves toward a global minimum. 
The lower plot illustrates the variation with $f$ of the local minimum corresponding to the 
true value of \al~for $f=0$. 

In view of the non-trivial shape of $\chi^2(\alpha)$ on figure ~\ref{fig:scan}, 
one should not express the measurement of \al~ in term of a central value and a
statistical error derived from $\Delta\chi^2(\alpha)=1$. Instead, one should rather provide
confidence levels as a function of \al~\cite{Sophie}.
Notwithstanding the above remark, half of the range defined by 
$\Delta\chi^2(\alpha)\leq 1$ leads to $\sigma(\alpha) = 0.23$ rad. This value should be compared 
to the systematic effect induced by the non-factorizable contributions: 
for, \eg, $f=0.1$, the bias in \al~is $0.18$ rad, which is comparable to the
statistical error.

\section{Other Charmless $B$ Decays related to \mbox{SCCFT}}

\subsection{Non-Resonant $B\to \eta\pi\pi$ Decay}

The non-resonant $B\to \eta\pi\pi$ decay is affected by the absence of the second-class current as well: 
the coupling $W \to \eta\pi$ remains forbidden since the $\eta\pi$ state is always produced 
with a natural spin-parity. As for \ap, this can lead to an enhancement of direct CP violation. 

Since the spin-parities of $\eta'(958)$ and $\eta(550)$ are identical,
both $B^0\to\eta\pi^+\pi^-$ and $B^0\to\eta'(958)\pi^+\pi^-$ decays should be considered. 
Contributions from channels like $B^0\to\eta(\eta')\rho^0$ contaminate the non-resonant 
signal sample, and have to be vetoed. 

\subsection{\ap~\vs $a_0 K$}

As in \pp\, the measurement of the ratio of ${\cal B}(B^0 \rar a_0\pi)/{\cal B}(B^0 \rar a_0 K)$, 
under some assumptions (\eg, neglecting the Cabibbo suppressed tree contribution in the $B^0\to a_0 K$ decay), can help to 
estimate the ratio of tree to penguin contributions to the \ap~decay. 
It also gives a handle on the charming penguin contributions.

\subsection{Analysis of $B^0\to b_1\pi$}

The $b_1$ resonance, with even $G$-parity and odd spin-parity, has the same properties 
leading to \mbox{SCCFT} as the $a_0$, so that the two-body analysis for \al~ can be performed accordingly.

Since the reconstruction of the $b_1$ proceeds through the decays $b_1\to\omega\pi\to 3\pi^\pm \pi^0$, 
the higher multiplicity of the final state and the lower energy of the $\pi^0$ renders
this mode less accessible. In addition, feed-through from $W\to\omega\pi$ from the $J^P = 1^-$ 
channel contaminates the $b_1(\omega\pi)\pi$ signal. 
On the other hand, the narrow $b_1$ and $\omega$ resonances and the helicity
distribution improve the background suppression.

Finally, the non-resonant $W \to \omega\pi$ transition can be produced in a $G$-parity allowed state due to the 
spin $1$ of the $\omega$. Therefore, direct CP searches in the non-resonant 
$B \rar \omega\pi\pi$ do not benefit from the absence of second class currents.

\subsection{Pure Penguin $a_0 a_0$, $b_1 b_1$ and $a_0 b_1$ Decays}
\label{sec:aa-bb}

Due to the absence of Second Class Currents, the decays $B^0\to a_0 a_0$, $B^0\to b_1 b_1$ and $B^0\to a_0 b_1$ (to both charged 
and neutral final states, the latter being Fierz-transformed of the former) proceed {\it via} penguins only. Therefore,
there should not be any direct CP violation in these decays, unless if other contributions carrying a 
different weak phase are present ($u$- and $c$-penguins, re-scattering from other final states). The observation
of direct CP violation in these decays thus provides a direct measurement of the non-factorizable contributions.

Similarly, the corresponding charged $B$ tree decays (including the color-suppressed ones, due to Fierz-transformation) 
are suppressed by both the absence of Second Class Currents and isospin conservation (Eq.~\ref{isosp1}). 
The gluonic ($u$-, $c$- and $t$-) penguin contributions to $B^{\pm}\to a_0^{\pm} a_0^0$ and $B^{\pm}\to b_1^{\pm} b_1^0$ 
are suppressed by isospin conservation when inserting the relation $P^{+-} = P^{-+}$ in Eq.~(\ref{isosp5}). Hence,
since both tree and gluonic penguin contributions are suppressed, the observation of the $B^{\pm}\to a_0^{\pm} a_0^0$ 
and $B^{\pm}\to b_1^{\pm} b_1^0$ decays provides again a measurement of the non-factorizable contributions.

Moreover, the time-dependent analysis of $B^0\to a_0 b_1$ allows the extraction of the strong phase difference 
between the two penguin amplitudes $P^{+-}$ and $P^{-+}$.
Nevertheless, since the $B\to a_0 b_1$ decay has one $\eta$ and four charged $\pi$ in the final
state, the extraction of the signal is marred by large combinatorial background.

\subsection{Decays into Higher Spin Mesons}

Due to angular momentum conservation, there is no coupling of virtual W to the hadronic states of  
spin larger than one. The corresponding tree diagrams do not contribute to the decay amplitude thus 
causing effects similar to those created by \mbox{SCCFT}. 

One example of such decays is $B^0 \to a_2(1320)\pi \to \eta\pi\pi$. Other
higher resonance excitations could be considered for similar analyses to those described in this article.

\section{Conclusion}

 Constraints imposed by the absence of second class currents  provide new opportunities for CP violation 
studies in charmless $B$ decays. In this article, we discussed how the CKM angle $\alpha$ can be extracted 
from analyses of $B$ decays into the final states $a_0\pi(\rho)$ 
in a more robust fashion than in the original isospin-pentagon analyses proposed for \rp and \pp. 
A similar analysis can be performed for the decays $b_1\pi$ and $\eta(\eta')\pi\pi$, but
these latter modes are experimentally more challenging.
Fits with four (if one theoretical amplitude or one ratio of amplitudes is added) to nine (with no such theoretical
input) parameters can be performed for each of these decays. A fit combining several channels
would reduce the number of mirror solutions, and decrease the error on \al.

Significant enhancement of direct CP asymmetries could arise in the following channels: \ap, \bp~and non-resonant 
$B\to \eta(\eta')\pi\pi$ due to the absence of second class currents, 
independently of the hypotheses needed for the extraction of \al~ (\ie, factorization and the neglect
of $u$- and $c$-penguins).

Finally, many of these decays can be used to test the factorization assumption, and measure the 
non-factorizable contributions. For a luminosity of the order of about $500 fb^{-1}$, the 
systematic bias on \al~, induced by non-factorizable contributions of the size of $10 \%$, 
remains of the same order than the statistic uncertainty.

\section*{Remark}

Factorization breaking can be studied in \ap~as described in this article, and
in a variety of other decays, following the idea that the suppression of factorizable
contributions allows to study the non-factorizable ones. CP-violation studies (measurement
of $2 \beta + \gamma$ and enhanced CP assymetries) can also been performed in these decays.
This has been independently described in two articles by M.~Diehl and G.~Hiller~\cite{Gudrun}.

\section*{Acknowledgements}

We are indebted to Roy Aleksan, Robert Cahn, Jerome Charles, Andreas H\"ocker and Francois Le Diberder 
for their contributions to this work, and for the fruitful {\it and} cheerful collaboration. 

This work was supported by the Lawrence Berkeley National Laboratory, USA, and the Laboratoire de 
l'Acc\'el\'erateur Lin\'eaire, France.


\begin{thebibliography}{99}

\bibitem{Quinn2} A.E. Snyder, H.R. Quinn, {\it Phys.Rev.D} {\bf 48} (1993) 2139
\bibitem{London} M. Gronau and D. London, {\it Phys.Rev.Lett.} {\bf 65} (1990) 3381
\bibitem{Quinn}  H.J. Lipkin, Y. Nir, H.R. Quinn, A.E. Snyder, {\it Phys.Rev.D} {\bf 44} (1991) 1454
\bibitem{Dighe} A.S. Dighe, C.S. Kim , {\it Phys.Rev.D} {\bf 62} (2000) 111302 
\bibitem{CLEO} A. Lyon for the CLEO collaboration, ``CP Violation Studies with Rare B Physics at CLEO'',
talk given at BCP4, Ise-Shime, Japan, Feb.19-23, 2001
\bibitem{Babar} T.J. Champion for the \babar~ collaboration, to be published in the proceedings of 
30th International Conference on High-Energy Physics (ICHEP 2000), Osaka, Japan, 27 Jul - 2 Aug 2000.
SLAC-PUB-8696, BABAR-PROC-00-13, {\it hep-ex/0011018}
\bibitem{Sophie} S.Versill\'e, {\em La violation de CP dans BaBar: \'etiquetage des 
m\'esons B et \'etude du canal $B\to 3\pi$}, PhD thesis (in French), Universit\'e de Paris Sud  (1999)
\bibitem{BABARa0pi}  The branching ratio of $B0 \rightarrow a_0^{\pm}\pi^{\mp}$ 
has been recently measured by the BABAR collaboration: \\
B.~Aubert {\it et al}, ``Search for $B0\rightarrow a_0(980)\pi$'',
July 2001, {\it hep-ex/0107075}
\bibitem{PDG} Particle Data Group, C.Caso {\it et al}, {\it Eur.Phys.J}~{\bf C3} (2000) 1
\bibitem{Jerome} J.~Charles, {\it Phys.Rev.D} {\bf59} (1999) 054007
\bibitem{BabarBook} \babar~ Collaboration, {\it The \babar~ Physics Book} (1998)
\bibitem{Gudrun} M.~Diehl and G.~Hiller, SLAC-PUB-8822, DESY-01-060, {\it hep-ph/0105194}
Published in {\it JHEP} 0106:067,2001 \\
M.~Diehl and G.~Hiller, SLAC-PUB-8837, DESY-01-061, {\it hep-ph/0105213}


\end{thebibliography}
\end{document}